\documentclass[aps,prd,twocolumn,showpacs,showkeys,nofootinbib,floatfix]{revtex4-1}

\usepackage{amssymb}
\usepackage{amsmath}
\usepackage{graphicx}
\usepackage{hyperref}

\begin{document}

\title{Constraints to Holographic Dark Energy Model via Type Ia Supernovae, Baryon Acoustic Oscillation and WMAP7}

\author{Lixin Xu}
\email{lxxu@dlut.edu.cn}

\affiliation{Institute of Theoretical Physics, School of Physics \&
Optoelectronic Technology, Dalian University of Technology, Dalian,
116024, P. R. China}

\begin{abstract}
In this paper, the holographic dark energy (HDE) model, where the future event horizon is taken as an IR cut-off, is confronted by using currently available cosmic observational data sets which include type Ia supernovae, baryon acoustic oscillation and cosmic microwave background radiation from full information of WMAP-7yr. Via the Markov Chain Monte Carlo method, we obtain the values of model parameter $c= 0.696_{-  0.0737-   0.132-  0.190}^{+   0.0736+   0.159+  0.264}$ with $1,2,3\sigma$ regions. Therefore one can conclude that at lest $3\sigma$ level the future Universe will be dominated by phantom like dark energy. It is not consistent with positive energy condition, however this condition must be satisfied to derive the holographic bound. It implies that the current cosmic observational data points disfavor the HDE model.

\end{abstract}



\maketitle

\section{Introduction}

In 1998, two supernovae teams discovered that our Universe is undergoing an accelerated expansion \cite{ref:Riess98,ref:Perlmuter99}. From that time on, a flood of models have been presented to explain the accelerated expansion phase. For the reviews about the accelerated expansion and dark energy, please see   
\cite{ref:DEReview1,ref:DEReview2,ref:DEReview3,ref:DEReview4,ref:DEReview5,ref:DEReview6,ref:DEReview7}.  
A natural candidate to dark energy is the cosmological constant (CC) which was first introduced by Einstein to realize a static universe about a century ago. In the context of quantum field
theory (QFT), CC has relations with the vacuum or zero point energy
density of quantum fields, via
\begin{equation}
\rho_{\Lambda}=\frac{1}{2}\int^{\Lambda}_{0}\frac{4\pi k^2
dk}{(2\pi)^3}\sqrt{k^2+m^2}\approx\frac{\Lambda^4}{16\pi^2},
\end{equation}
where $\Lambda\gg m$ is a UV cut-off. To balance an assumed UV
cut-off $\Lambda$ and the observational smallness of CC, tremendous
fine tuning is required. This is the so-called cosmological constant
problem \cite{ref:DEReview1}. QFT in curved space-time leads to an infinite effective action or
vacuum expectation values (VEV) of the energy-momentum tensors of
the fields. A renormalization treatment can yield a scale-dependent
or running CC and a running Newton constant. Though, the absolute
values can not be calculated, the change with respect to the
renormalization scale can be calculated via renormalization group equations (RGEs) originating in QFT
\cite{ref:RGEsQFT} and quantum gravity \cite{ref:RGEsQG}. In the
cosmological context, it is reasonable to identify the
renormalization scale with some characteristic scales of cosmology.
The related investigation can be found in \cite{ref:PhDthesis},
where the renormalization scale $\mu$ was given by the Hubble scale
$H$, the inverse radius $R^{-1}$ of the cosmological event horizon
and the inverse radius $T^{-1}$ of the particle horizon. 

Another interesting viewpoints about the relations between vacuum energy density and a cosmological scale come from the so-called holographic principle in cosmology. According to the holographic principle, the number of degrees of freedom in a bounded system should be finite and has relations with the area of its boundary. By applying the principle to cosmology, one can obtain
the upper bound of the entropy contained in the universe. For a
system with size $L$ and UV cut-off $\Lambda$ without decaying into
a black hole, it is required that the total energy in a region of
size $L$ should not exceed the mass of a black hole of the same
size, thus $L^3\rho_{\Lambda} \le L M^2_{pl}$. The largest $L$
allowed is the one saturating this inequality, thus $\rho_{\Lambda}
=3c^2 M^{2}_{pl} L^{-2}$, where $c$ is a numerical constant and
$M_{pl}$ is the reduced Planck Mass $M^{-2}_{pl}=8 \pi G$.
It just means a {\it duality} between UV cut-off and IR cut-off. The
UV cut-off is related to the vacuum energy, and IR cut-off is
related to the large scale of the universe, for example Hubble
horizon, event horizon or particle horizon as discussed by
\cite{ref:holo0,ref:holo1,ref:holo2}. Then in this context the so-called fine tuning problem is removed naturally.

In the paper \cite{ref:holo2}, the
author proposed a dark energy model, dubbed holographic dark energy (HDE), where the future event horizon 
\begin{equation}
R_{eh}(a)=a\int^{\infty}_{t}\frac{dt^{'}}{a(t^{'})}=a\int^{\infty}_{a}\frac{da^{'}}{Ha^{'2}}\label{eq:EH}
\end{equation}
was taken as the IR cut-off $L$. This horizon is the boundary of the volume a
fixed observer may eventually observe. One is to formulate a theory
regarding a fixed observer within this horizon. As pointed out in
\cite{ref:holo2}, it can reveal the dynamic nature of the vacuum
energy and provide a solution to the {\it fine tuning} and {\it
cosmic coincidence} problem. 

The concrete value of parameter $c$ is very important to determine the properties of HDE. For example in the epoch when the HDE becomes to dominate the energy density, i.e. $\Omega_{h}\sim 1$, if $c\ge 1$, $c=1$ and $c\le 1$ are respected, the HDE behaves like quintessence, cosmological constant and phantom respectively. One can clearly see that in the following Eq. (\ref{eq:DEEOS}) by setting $\Omega_{h}=1$. In the past years, the HDE model was confronted with cosmic observations. In Ref. \cite{ref:Kao2005}, the authors used the angular scale of the acoustic oscillation from the BOOMRANGE and WMAP data on CMB to constrain HDE model. Also the authors of Ref. \cite{ref:Gong2005} used the CMB shift parameter $R$ to constrain HDE model. For the recent result please see \cite{ref:holoZhang} where the best fit value of parameter $c=0.65^{+0.10}_{-0.08}$ in $1\sigma$ region was given. In that paper, about the CMB information only the CMB shift parameter $R$, acoustic scale $l_A$ and the redshift $z_\ast$ at the decoupling epoch of photons obtained from WMAP-7yr in $\Lambda$CDM model were employed. The main problem comes from the fact that the values of the CMB shift parameters used in the literature for example \cite{ref:holoZhang}, depend on the cosmological model, here the $\Lambda$CDM model. So, when one uses these derived data points to constrain other cosmological models, the so-called circular problem, i.e. dependence on cosmological models, would be committed. And the circular problem is what we try to avoid in the cosmological constraint issue. Therefore one should constrain the cosmological model in a consistent way by using cosmic observational data points which do not depend on any cosmological model. However to use full CMB data sets from WMAP7-yr to constrain the HDE model is still missing in the literature. In this paper to fill up the gap, we will use the full information from WMAP-7yr data sets in stead of CMB shift parameters. Furthermore, the CMB shift parameters contain information around the last scattering surface of background photons where the contribution from dark energy may not be neglected as CC case due to a nontrivial equation of state (EoS) of dark energy, for example early dark energy model \cite{ref:EDE}. In this situation, it is dangerous to use theses derived data points from $\Lambda$CDM model to constrain other cosmological model due to much departure from $\Lambda$CDM model. At late epoch when the dark energy dominates, the gravitational potentials decay and the late integrated Sachs-Wolfe (ISW) arises. It affects to the anisotropy power spectra of CMB at large scale (low $l$ parts). The ISW effect is sensitive to the properties of dark energy. Apparently, the late ISW effect is not expressed by the CMB shift parameters. So one can expect a tight constraint to cosmological model parameter space when the late ISW effect is included. 

Base on these points mentioned above, in this paper we will not use the derived CMB shift parameters but the full information from WMAP-7yr data points to constrain the HDE model. The important is that there doesn't exist any circular problem. Due to the full information from WMAP-7yr is used, one can expect to obtain a tight constraint. At low redshift regions, the type Ia supernova (SN Ia) and baryon acoustic oscillations (BAO) are also included as geometric measures.

This paper is structured as follows. In section \ref{sec:bgpe}, we give a brief review of the HDE model where the radiation is included and the future event horizon adopted as an IR cut-off. The scalar perturbation evolution equations for a spatially flat FRW Universe will also be presented. In section \ref{sec:method}, the constraint methodology and results will be presented. We give a summary in section \ref{sec:summary}.
 
\section{Background and Perturbation Evolution Equations} \label{sec:bgpe}

Now, as is done in \cite{ref:holo2}, the event horizon $R_{eh}$ 
taken as the IR cut-off. Then, the HDE is written as
\begin{equation}
\rho_{h}=\frac{3c^2 M^2_{pl}}{R^2_{eh}},\label{eq:EHHDE}
\end{equation}
and the Friedmann equation in a spatially flat FRW Universe can be written as
\begin{equation}
H^2=H^2_0\left(\Omega_{r0}a^{-4}+\Omega_{b0}a^{-3}+\Omega_{c0}a^{-3}\right)+\Omega_{h}H^2,\label{eq:EHFE}
\end{equation}
where $\Omega_{i}=\frac{\rho_i}{3 M^2_{pl}H^2}$ are dimensionless energy densities for radiation, baryon, cold dark matter and HDE respectively. For convenience, the scale factor $a$ has been
normalized to $a_0=1$. Combining Eq. (\ref{eq:EHHDE}) and Eq.
(\ref{eq:EH}), one has
\begin{equation}
\int^{\infty}_{a}\frac{d\ln
a'}{Ha'}=\frac{c}{aH\sqrt{\Omega_{h}}},\label{eq:re}
\end{equation}
where the definition of dimensionless energy density parameter for HDE is used.
From Eq.(\ref{eq:EHFE}), one obtains
\begin{equation}
\frac{1}{Ha}=\frac{\sqrt{1-\Omega_{h}}}{a H_0
\sqrt{\left(\Omega_{r0}a^{-4}+\Omega_{b0}a^{-3}+\Omega_{c0}a^{-3}\right)}}.
\end{equation}
Inserting the above equation into Eq. (\ref{eq:re}), one has
\begin{equation}
\int^{\infty}_{x}\frac{\sqrt{1-\Omega_{h}}}{E(x')}dx'=\frac{c\sqrt{1-\Omega_{h}}}{E(x)\sqrt{\Omega_{h}}},\label{eq:difomehah}
\end{equation}
where $E(x)=\sqrt{\Omega_{r0}e^{-2x}+\Omega_{b0}e^{-x}+\Omega_{c0}e^{-x}}$ and $x=\ln a$. Taking derivative with respect to $x =\ln a$ from
both sides of the above equation, one has the differential equation
of $\Omega_{h}$
\begin{equation}
\Omega_{h}'=-2\Omega_{h}\left(1-\Omega_{h}\right)\left(\frac{E'(x)}{E(x)}-\frac{\sqrt{\Omega_{h}}}{c}\right),\label{eq:diffeq}
\end{equation}
where $'$ denotes the derivative with respect to $x =\ln a$. This
equation with initial condition $\Omega_{h0}=1-\Omega_{r0}-\Omega_{b0}-\Omega_{c0}$ describes the evolution of dimensionless energy density of
dark energy. Then the background evolution will be obtained by solving this differential equation and Friedmann equation. When the baryon and radiation components are omitted, one has $E'(x)/E(x)=-1/2$. Then the evolution equation for $\Omega_h$ \cite{ref:holo2}
\begin{equation}
\Omega_{h}'=\Omega_{h}\left(1-\Omega_{h}\right)\left(1+\frac{2}{c}\sqrt{\Omega_{h}}\right),
\end{equation}
is recovered. From the
conservation equation of the HDE
$\dot{\rho_{h}}+3H(\rho_{h}+p_{h})=0$, one has the equation of state (EoS) of holographic
dark energy
\begin{equation}
w_{h}=-1-\frac{1}{3}\frac{d \ln \rho_{h}}{d \ln a}
%=-1-\frac{1}{3}\left(2\frac{d\ln H}{d\ln a}+\frac{d\ln \Omega_h}{d\ln a}\right)
=-\frac{1}{3}-\frac{2\sqrt{\Omega_{h}}}{3c}\label{eq:DEEOS}
\end{equation}
where $w_{h}=p_{h}/\rho_{h}$.

In this paper, we take the HDE as a perfect fluid having EoS (\ref{eq:DEEOS}). Considering the perturbation in the synchronous gauge and the conservation of energy-momentum tensor $T^{\mu}_{\nu;\mu}=0$, one has the perturbation equations of density contrast and velocity divergence for the HDE
\begin{eqnarray}
\dot{\delta}_h=-(1+w_h)(\theta_d+\frac{\dot{h}}{2})-3\mathcal{H}(c^{2}_{s}-w_h)\delta_h\\
\dot{\theta}_h=-\mathcal{H}(1-3c^{2}_{s})\theta_h+\frac{c^{2}_{s}}{1+w_h}k^{2}\delta_h-k^{2}\sigma_h
\end{eqnarray}
following the notations of Ma and Bertschinger \cite{ref:MB}. For the gauge ready formalism about the perturbation theory, please see \cite{ref:Hwang}. For the HDE, we assume the shear perturbation $\sigma_h=0$. 
In our calculations, the adiabatic initial conditions will be taken.

\section{Methodology and Constraint Results} \label{sec:method}

Before constraining the model parameter space, we study the effects of model parameter $c$ to the CMB power spectrum by setting different values of $c$ and fixing the other relevant cosmological model parameters. At fist, we modify the CAMB package \cite{ref:CAMB}, which is the code for calculating the CMB power spectrum and is publicly available, to include the HDE. The differential equation (\ref{eq:diffeq}) was solved by taking $\Omega_{h0}=1-\Omega_{m0}-\Omega_{r0}$ as initial condition. We borrow and fix the other relevant cosmological parameters values from WMAP7, the resultant CMB power spectra are shown in Fig. \ref{fig:cls} for different values of $c$, where $h=0.72$,  $\omega_c=0.112$, $\omega_b=0.0226$, $n_s=0.96$ are fixed.
\begin{widetext}
\begin{center}
\begin{figure}[htb]
\includegraphics[width=14cm]{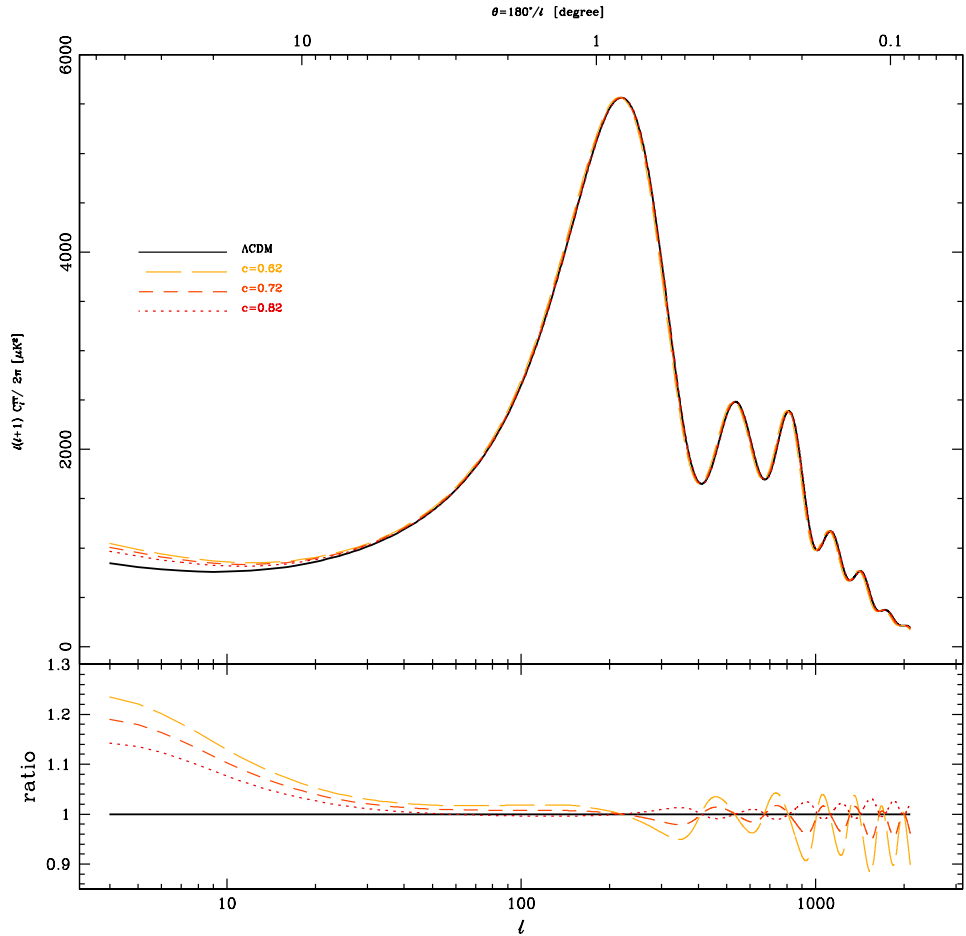}
\caption{The CMB $C^{TT}_l$ power spectrum v.s. multiple moment $l$ and the corresponding ratio to $\Lambda$CDM model for different values of model parameter $c$, where $h=0.72$,  $\omega_c=0.112$, $\omega_b=0.0226$, $n_s=0.96$ are fixed. The main difference appears at low ($l<20$) multipole momentum parts which correspond to large scale.}\label{fig:cls}
\end{figure}
\end{center}
\end{widetext}

From Fig. \ref{fig:cls}, one can see that the main contribution to CMB power spectra from the HDE comes from the evolution of EoS and its domination at the late epoch through late ISW effect. At early epoch the contribution to CMB power spectra is small due to small energy density ratio of the HDE, that can be understood from the EoS of the HDE (\ref{eq:DEEOS}) which seems close to $-1/3$ before the last scattering surface. It means that a tight constraint would be obtained when CMB information from large scale is included. But unfortunately, the CMB data points are dominated by the so-called cosmic variance at large scale, i.e. low $l$ part. 

We perform a global fitting to the model parameter space by using the Markov Chain Monte Carlo (MCMC) method. We modified publicly available {\bf cosmoMC} package \cite{ref:MCMC} to include the HDE in the CAMB \cite{ref:CAMB} code which is used to calculate the theoretical CMB power spectrum. The following $7$-dimensional parameter space  is adopted
\begin{equation}
P\equiv\{\omega_{b},\omega_c, \Theta_{S},\tau, c,n_{s},\log[10^{10}A_{s}]\}
\end{equation}
where $\omega_{b}=\Omega_{b}h^{2}$ and $\omega_{c}=\Omega_{c}h^{2}$ are the physical density of  baryon and cold dark matter respectively, $\Theta_{S}$ (multiplied by $100$) is the ration of the sound horizon and angular diameter distance, $\tau$ is the optical depth, $c$ is the newly added model parameter related to HDE, $n_{s}$ is scalar spectral index, $A_{s}$ is the amplitude of of the initial power spectrum. The pivot scale of the initial scalar power spectrum $k_{s0}=0.05\text{Mpc}^{-1}$ is used in this paper. The following priors to model parameters are adopted: $\omega_{b}\in[0.005,0.1]$, $\omega_{c}\in[0.01,0.99]$, $\Theta_{S}\in[0.5,10]$, $\tau\in[0.01,0.8]$, $c\in[0,1]$, $n_{s}\in[0.5,1.5]$, $\log[10^{10}A_{s}]\in[2.7, 4]$. Furthermore, the hard coded prior on the comic age $10\text{Gyr}<t_{0}<\text{20Gyr}$ is also imposed. Also, the physical baryon density $\omega_{b}=0.022\pm0.002$ \cite{ref:bbn} from big bang nucleosynthesis and new Hubble constant $H_{0}=74.2\pm3.6\text{kms}^{-1}\text{Mpc}^{-1}$ \cite{ref:hubble} are adopted.

To get the distribution of parameters, we calculate the total likelihood $\mathcal{L} \propto e^{-\chi^{2}/2}$, where $\chi^{2}$ is given as
\begin{equation}
\chi^{2}=\chi^{2}_{CMB}+\chi^{2}_{BAO}+\chi^{2}_{SN}.
\end{equation}
The $557$ Union2 data \cite{ref:Union2} with systematic errors and BAO \cite{ref:BAO} are used to constrain the background evolution, for the detailed description please see Refs. \cite{ref:Xu}. SN Ia is used as standard candle. And BAO is used as standard ruler. At the last scattering of CMB radiation, the acoustic oscillation in the baryon-photon fluid was frozen and imprinted their signature on the matter distribution. The characterized scale of BAO in the observed galaxy power spectrum is determined by the comoving sound horizon at drag epoch $z_d$ which is shortly after photon decoupling and defined as the redshift at which the drag optical depth $\tau_d$ equals one. To calculate $r_{s}(z_{d})$, one needs to know the redshift $z_d$ at the drag epoch and its corresponding sound horizon. The baryon drag epoch redshift $z_d$ can be obtained from its definition and be calculated numerically \cite{ref:Hamann}
\begin{eqnarray}
\tau(\eta_d)&\equiv& \int_{\eta}^{\eta_0}d\eta'\dot{\tau}_d\nonumber\\
&=&\int_0^{z_d}dz\frac{d\eta}{da}\frac{x_e(z)\sigma_T}{R}=1
\end{eqnarray}   
where $R=3\rho_{b}/4\rho_{\gamma}$, $\sigma_T$ is the Thomson cross-section and $x_e(z)$ is the fraction of free electrons. Then the sound horizon is
\begin{equation}
r_{s}(z_{d})=\int_{0}^{\eta(z_{d})}d\eta c_{s}(1+z).
\end{equation}   
where $c_s=1/\sqrt{3(1+R)}$ is the sound speed. Also, to obtain unbiased parameter and error estimates, we use the substitution \cite{ref:Hamann}
\begin{equation}
d_z\rightarrow d_z\frac{\hat{r}_s(\tilde{z}_d)}{\hat{r}_s(z_d)}r_s(z_d),
\end{equation}
where $d_z=r_s(\tilde{z}_d)/D_V(z)$, $\hat{r}_s$ is evaluated for the fiducial cosmology of Ref. \cite{ref:BAO}, and $\tilde{z}_d$ is redshift of drag epoch obtained by using the fitting formula \cite{ref:EH} for the fiducial cosmology. Here $D_V(z)=[(1+z)^2D^2_Acz/H(z)]^{1/3}$ is the 'volume distance' with the angular diameter distance $D_A$. In this paper, for BAO information, the SDSS data points from \cite{ref:BAO} and the WiggleZ data points \cite{ref:wiggles} are used. For CMB data set, the temperature power spectrum from WMAP $7$-year data \cite{ref:wmap7} are employed as dynamic constraint.

After running $8$ independent chains and checking the convergence to stop sampling when the worst e-values [the variance(mean)/mean(variance) of 1/2 chains] $R-1$ is of the order $0.01$, the global fitting results are summarized in Table \ref{tab:results} and Figure \ref{fig:contour}. We find that the values of $c$ in $1,2,3\sigma$ regions are $c= 0.696_{-  0.0737-   0.132-  0.190}^{+   0.0736+   0.159+  0.264}$, which are compatible with the result obtained in \cite{ref:holoZhang}. And the error bars are slightly smaller than that in \cite{ref:holoZhang}. This result shows that the values of $c$ are less than $1$ at $3\sigma$ level. Correspondingly, we plot the CMB power spectrum for the mean values estimated from MCMC analysis in Figure \ref{fig:mean}, where the observational data points of $7$-year WMAP with uncertainties are also included. As comparisons, the CMB temperature power spectrum for $\Lambda$CDM with same cosmic observational data sets combination and results from \cite{ref:wmap7} are also shown. One can see that the CMB power spectrum for the HDE is well inside the error bars of the binned measurements from WMAP $7$-year results and matches to $\Lambda$CDM model very well. This strongly implies that current cosmic observational data combinations of SN Union2, BAO and WMAP $7$-year can not almost discriminate the HDE  from $\Lambda$CDM model. In this sense, the HDE is a competitive model of dark energy.

\begingroup
%\squeezetable
\begin{table}
\begin{center}
\begin{tabular}{ccc}
\hline\hline Prameters&Mean with errors & Best fit \\ \hline
$\Omega_b h^2$ & $    0.0226_{-    0.000546-    0.00107}^{+    0.000536+    0.00108}$ & $0.0226$\\
$\Omega_{DM} h^2$ & $    0.110_{-    0.00414-    0.00838}^{+    0.00422+    0.00866}$ & $0.112$\\
$\theta$ & $    1.0396_{-    0.00263-    0.00520}^{+    0.00268+    0.00513}$ & $1.0400$\\
$\tau$ & $    0.0892_{-    0.00719-    0.0226}^{+    0.00635+    0.0251}$ & $0.0865$\\
$c$ & $    0.696_{-    0.0737-    0.132}^{+    0.0736+    0.159}$ & $0.643$\\
$n_s$ & $    0.971_{-    0.0134-    0.0255}^{+    0.0135+    0.0262}$ & $0.971$\\
$\log[10^{10} A_s]$ & $    3.0802_{-    0.0337-    0.0646}^{+    0.0340+    0.0704}$ & $3.0805$\\
$\Omega_h$ & $    0.727_{-    0.0171-    0.0349}^{+    0.0172+    0.0323}$ & $0.733$\\
$Age/Gyr$ & $   13.878_{-    0.113-    0.216}^{+    0.110+    0.221}$ & $13.839$\\
$\Omega_m$ & $    0.273_{-    0.0172-    0.0323}^{+    0.0171+    0.0349}$ & $0.267$\\
$z_{re}$ & $   10.623_{-    1.161-    2.229}^{+    1.189+    2.352}$ & $10.468$\\
$H_0$ & $   69.916_{-    1.844-    3.585}^{+    1.867+    3.635}$ & $71.0657$\\
\hline\hline
\end{tabular}
\caption{The mean values of model parameters with $1\sigma$ and $2\sigma$ errors, where WMAP $7$-year, SN Union2 and BAO data sets are used.}\label{tab:results}
\end{center}
\end{table}
\endgroup

\begin{widetext}
\begin{center}
\begin{figure}[htb]
\includegraphics[width=17cm]{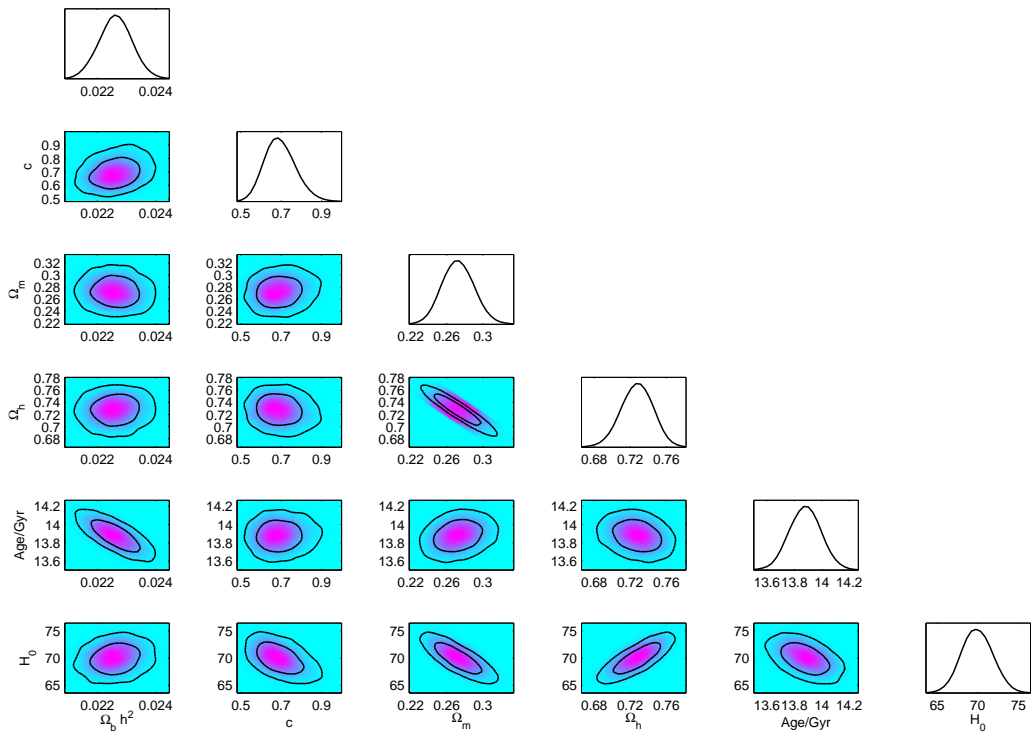}
\caption{The 1D marginalized distribution on individual parameters and 2D contours  with $68\%$ C.L. and $95\%$ C.L. by using CMB+BAO+SN data points. The shade regions show the mean likelihood of the samples.}\label{fig:contour}
\end{figure}
\end{center}
\end{widetext}
\begin{widetext}
\begin{center}
\begin{figure}[htb]
\includegraphics[width=17cm]{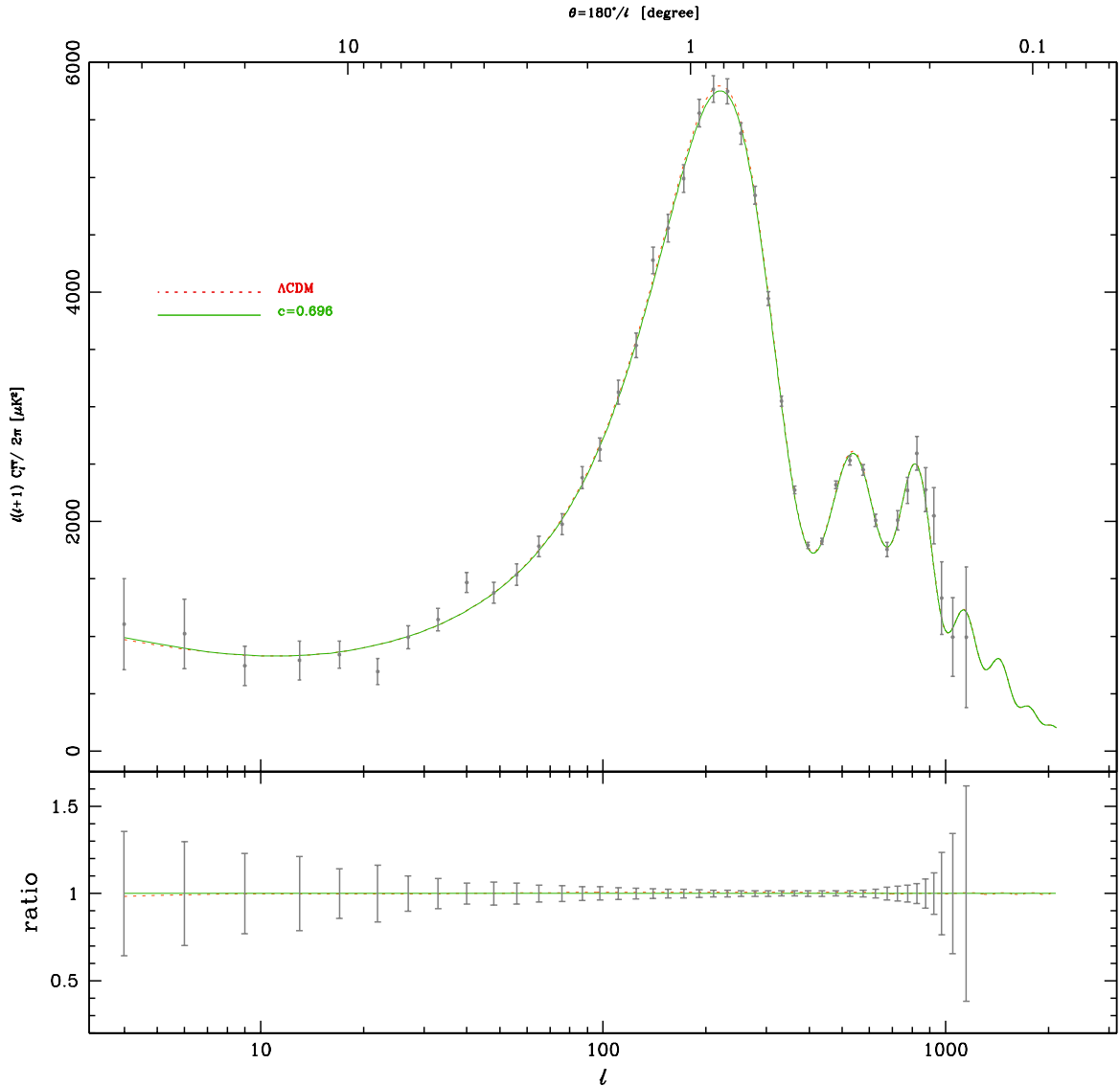}
\caption{The CMB $C^{TT}_l$ power spectrum v.s. multiple moment $l$, where the grey dots with error
bars denote the observed data with their corresponding uncertainties from WMAP $7$-year results, the green solid line is for HDE with mean values as shown in Table \ref{tab:results}, the red dotted line is for $\Lambda$CDM model with mean values taken from \cite{ref:wmap7} with WMAP+BAO+$H_0$ constraint results.}\label{fig:mean}
\end{figure}
\end{center}
\end{widetext}

We can show the evolutions of dimensionless energy density of the HDE and the EoS of the HDE with respect to redshift $z$ in Fig. \ref{fig:omhwh}, one can solve the differential equation (\ref{eq:difomehah}) by setting the initial conditions $\Omega_{h0}$ and model parameters $\Omega_{m0}=\Omega_{b0}+\Omega_{c0}$ and $c$ to their central values as listed in Table \ref{tab:results}, here $\Omega_{r0}$ is derived from the temperature of CMB background $T_0=2.726 K$. 
\begin{widetext}
\begin{center}
\begin{figure}[htb]
\includegraphics[width=8cm]{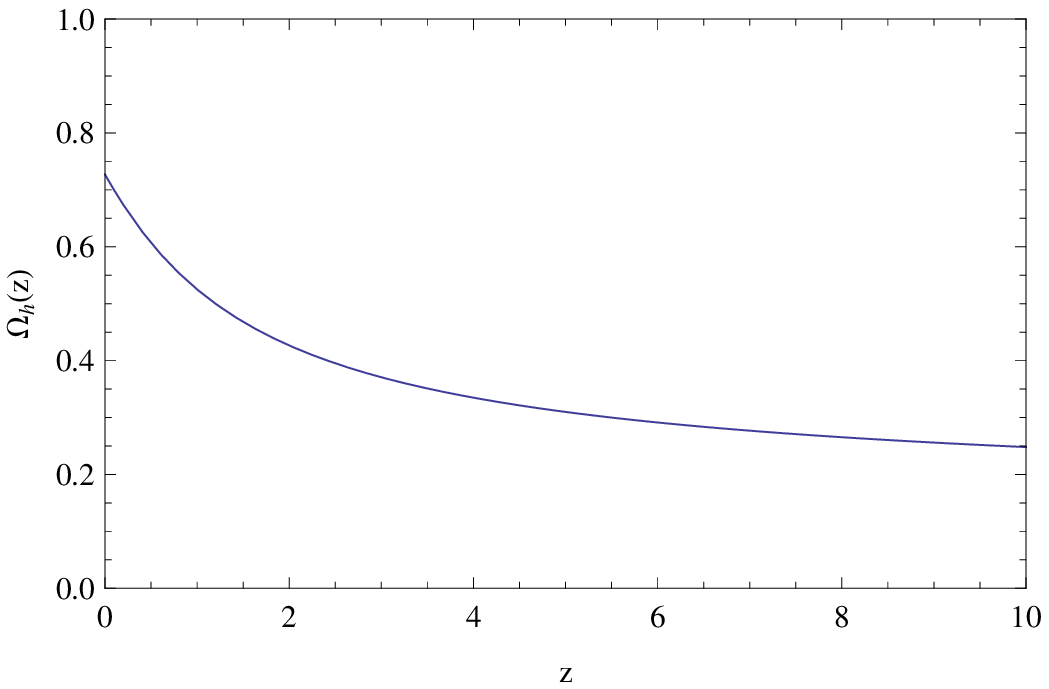}
\includegraphics[width=8cm]{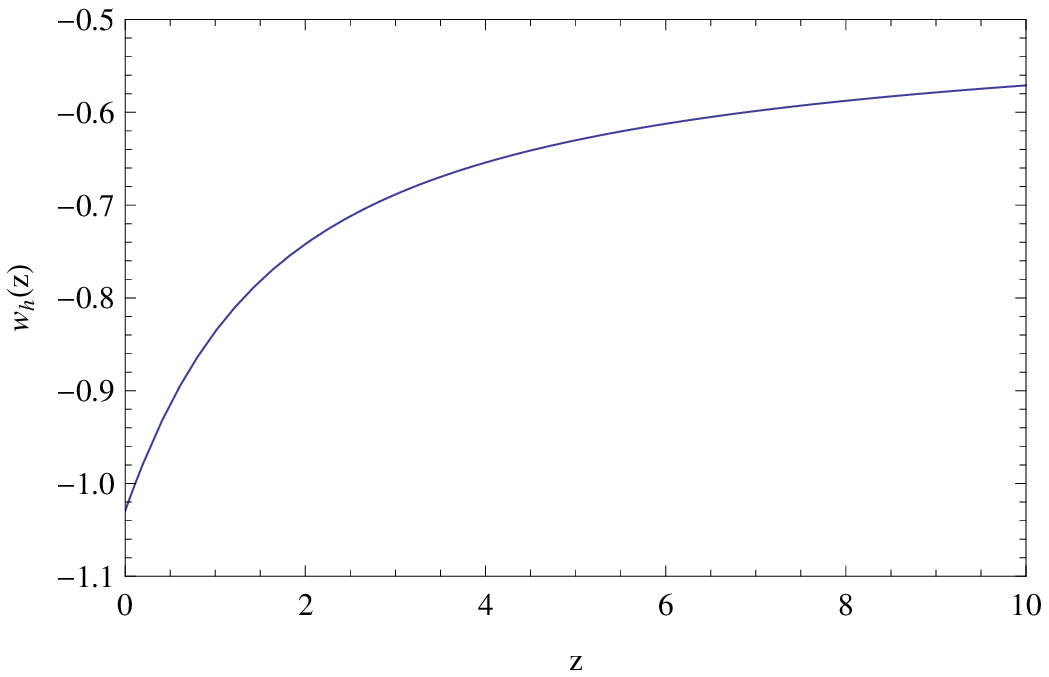}
\caption{The evolutions of dimensionless energy density of the HDE (left panel) and EoS of the HDE (right panel) with respect to redshift $z$, where the central values of model parameters are taken as listed in Table \ref{tab:results}.}\label{fig:omhwh}
\end{figure}
\end{center}
\end{widetext}
From the Fig. \ref{fig:omhwh}, one can read off that the HDE is phantom like at late time and quintessence like at early epoch.

\section{Summary and Discussion} \label{sec:summary}

In this paper, we perform a global fitting to the HDE model by using MCMC method with the combination of the full information from CMB, BAO and  SN Ia data points. We analyze the effect of model parameter $c$ to the CMB power spectra by taking its different values. We find out that $c$ mainly contributes to CMB power spectra at large scale where the so-called cosmic variance is dominated. However, we still obtain a relative tight constrain to model parameter space where the values of $c$ with $1,2,3\sigma$ regions are achieved: $c= 0.696_{-  0.0737-   0.132-  0.190}^{+   0.0736+   0.159+  0.264}$. So at $3\sigma$ level, $c$ is a number less than $1$. As a comparison to the previous results obtained in \cite{ref:holoZhang}, a slightly tight constraint is achieved in this paper. More important is that a consistent constraint is carried out instead of using the derived CMB shift parameters. Based on this result, one can conclude that at least $3\sigma$ level the future Universe will be dominated by phantom like dark energy. Then the positive energy condition is not respected, however this condition must be satisfied to derive the holographic bound. So the current cosmic observational data points disfavor the HDE model based on our analysis.   

Based on the holographic principle,  other HDE models were proposed in the literature where different IR cut-offs were adopted. For example, the Hubble horizon as IR cut-off was discussed when the model parameter $c$ is constant \cite{ref:holo1,ref:holo2} or time variable \cite{ref:XuJCAP}. In Ref. \cite{ref:Gao}, the authors took the Ricci scalar as the IR cut-off and named it Ricci dark energy, the corresponding cosmic observational constraint to this model could be found in Ref. \cite{ref:WangXu}. Interestingly, Cai, {\it et. al.} found out that the holographic Ricci dark energy had relations with the causal connection scale $R^{-2}_{CC} = \text{Max}(\dot{H} + 2H^2,-\dot{H} )$ for a spatially flat Universe \cite{ref:CAI}. Based on the idea that gravity as an entropic force \cite{ref:Verlinde},  a similar DE density was given in \cite{ref:EFS} where a linear combination of $H^2$ and $\dot{H}$ was also presented, see also \cite{ref:Hcom1,ref:Hcom2}. And a generalized HDE model $\rho_{R}=3c^2M^{2}_{pl}Rf(H^2/R)$ and $\rho_{h}=3c^2M^{2}_{pl}H^2g(R/H^2)$ were presented in Ref. \cite{ref:XuGHDE}. To the best of our knowledge, the full information from CMB has not been used to test these models yet. One can apply the method and data sets presented in this paper to these models directly. 

\acknowledgements{The author thanks an anonymous referee for helpful improvement of this paper and Dr. Chan-Gyung Park for helping to plot the figures for CMB power spectra. L. Xu's work is supported by the Fundamental Research Funds for the Central Universities (DUT10LK31) and (DUT11LK39).}

\end{document}